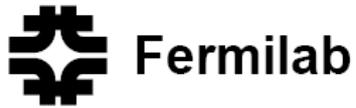

# TEVATRON BEAM HALO COLLIMATION SYSTEM: DESIGN, OPERATIONAL EXPERIENCE AND NEW METHODS*†

Nikolai Mokhov, Jerry Annala, Richard Carrigan, Michael Church, Alexander Drozhdin, Todd Johnson, Robert Reilly, Vladimir Shiltsev, Giulio Stancari[+], Dean Still, Alexander Valishev, Xiao-Long Zhang, Viktoriya Zvoda

*Fermilab, P.O. Box 500*
*Batavia, IL 60510, U.S.A.*

[*]Work supported by Fermi Research Alliance, LLC under contract No. DE-AC02-07CH11359 with the U.S. Department of Energy.

[†]To be published in *Journal of Instrumentation (JINST)*.

[+]On leave from Istituto Nazionale di Fisica Nucleare (INFN), Sezione di Ferrara, Italy.

# Tevatron Beam Halo Collimation System: Design, Operational Experience and New Methods[1]

NIKOLAI MOKHOV[*], JERRY ANNALA, RICHARD CARRIGAN, MICHAEL CHURCH, ALEXANDER DROZHDIN, TODD JOHNSON, ROBERT REILLY, VLADIMIR SHILTSEV, GIULIO STANCARI[+], DEAN STILL, ALEXANDER VALISHEV, XIAO-LONG ZHANG AND VIKTORIYA ZVODA

*Fermi National Accelerator Laboratory*
*PO Box 500, Batavia, IL, 60510, USA*

**ABSTRACT:** Collimation of proton and antiproton beams in the Tevatron collider is required to protect CDF and D0 detectors and minimize their background rates, to keep irradiation of superconducting magnets under control, to maintain long-term operational reliability, and to reduce the impact of beam-induced radiation on the environment. In this article we briefly describe the design, practical implementation and performance of the collider collimation system, methods to control transverse and longitudinal beam halo and two novel collimation techniques tested in the Tevatron.

KEYWORDS: Accelerator Subsystems and Technologies; Instrumentation for particle accelerators and storage rings, high energy (linear accelerators, synchrotrons); bent crystal collimation; hollow electron beam collimation

[1] Work supported by Fermi Research Alliance, LLC under Contract No. DE-AC02-07CH11359 with the US Department of Energy
[*] Corresponding author: E-mail: `mokhov@fnal.gov`
[+] On leave from Istituto Nazionale di Fisica Nucleare (INFN), Sezione di Ferrara, Italy

# Contents

## 1. Tevatron Collider Operational Halo Removal Systems

**1.1 Introduction**

Even in good operational conditions, a finite fraction of the beam will leave the stable central area of an accelerator because of beam-gas interactions, intra-beam scattering, proton-antiproton interactions in the IPs, RF noise, ground motion and resonances excited by the accelerator elements imperfection. These particles form a beam halo. As a result of halo interactions with limiting apertures, hadronic and electromagnetic showers are induced in accelerator and detector components causing numerous deleterious effects ranging from minor to severe. The most critical for colliders are beam losses in superconducting magnets and accelerator related backgrounds in the collider detectors. Only with a very efficient beam collimation system can one reduce uncontrolled beam losses in the machine to an allowable level. Beam collimation is mandatory at any superconducting hadron collider to protect components against excessive irradiation, minimize backgrounds in the experiments, maintain operational reliability over the life of the machine, and reduce the impact of radiation on the environment [1].

During the Collider Run I (1994-1996) the Tevatron halo removal system experienced limitations that prompted a design of a new system for the Collider Run II. The new design specified that the entire halo removal process needed to be more efficient and conducted in approximately 5 min. This implied that the halo removal process would have to be based on a two-stage collimation [1] and move toward automation. A new collimation system [2] was designed for the Tevatron Run II to localize most of the losses in the straight sections D17, D49, EØ, F17, F48, F49 and AØ. It incorporated four *primary collimators* (targets) and eight newly built 1.5-m long *secondary collimators*. New motion control hardware capable of fast processing of beam loss monitor and beam intensity feedback control with motor speeds that would allow the 2 inch full travel of the collimator to take 15 sec were also specified. A central control software system was also developed to coordinate the global sequence of motion for all 12 collimators while incorporating the halo removal system into the Tevatron Collider sequencer software. At the design stage, a multi-turn particle tracking through the accelerator and beam halo interactions with the collimators was done with the STRUCT [3] code. Using the calculated beam loss distributions, Monte-Carlo hadronic and electromagnetic shower simulation, secondary particle transport in the accelerator and detector components, including shielding with real materials and magnetic fields were done with the MARS [4] code. The Collider Run II halo removal system was installed, commissioned and has now been operational since June 2001.

The system was upgraded several times following operational needs. For example, in 2002, the Tevatron Electron Lenses were set up to remove undesirable uncaptured particles from the abort beam gaps and, thus, reduce the risk of damage of high-energy physics particle detectors at CDF and D0 during beam aborts. In 2003, following several instances of unsynchronized abort kicker pre-fires in the Tevatron, an additional *tertiary* collimator was installed at the A48 location to protect the CDF detector components [5]. Beam scraping procedures were optimized for faster operation and highly-efficient repeated scraping (double scraping) of the beams was made operational in 2005. Also, in 2010 collimation during the low-beta squeeze was added in order to reduce losses at CDF and D0 that were causing frequent quenches. Quenching the low-beta quadrupoles during the squeeze became more of a problem once the antiproton intensity and beam brightness became larger. Sensitive steps in the low-beta squeeze, where the sigma separation between the proton and antiproton is small, create losses at large beta locations mainly the cryogenic low-beta quadrupoles. A single collimator at E0 is placed at 5 sigma to create a limiting aperture moving the loss point of sensitive steps away from CDF and D0 IP's to a region that has robust quench limits. This additional collimator has worked well allowing the number of antiprotons to be transmitted through the low-beta squeeze to be increased while limiting the number of quenches in the low-beta squeeze to 0 since it was employed. Novel ideas to improve beam collimation efficiency – namely, a bent crystal collimation and hollow electron beam collimation – have been extensively and successfully studied at the end of Run II.

**1.2 Collimation System Design**

The principles of a two-stage collimation system are described in Ref. [1]. The system consists of horizontal and vertical primary collimators and a set of secondary collimators placed at an optimal phase advance, to intercept most of the particles out-scattered from the primary collimators during the first turn after beam halo interaction with primary collimators. An impact parameter of multi-GeV and TeV protons on the primary collimators is ~1 μm [6]. The design studies [2] show that in the Tevatron, a 5-mm thick tungsten primary collimator positioned at $5\sigma$ (rms beam size) from the beam axis in both vertical and horizontal planes would function optimally, reducing the beam loss rates as much as a factor of 4 to 10 compared to the system without such a scatterer. Secondary collimators, located at the appropriate phase advances, are 1.5-m long L-shaped steel jaws positioned at 6 $\sigma$ from the beam axis in horizontal and vertical planes. They are aligned parallel to the envelope of the circulating beam. Figure 1 depicts schematically the placement of the collimators in such a system.

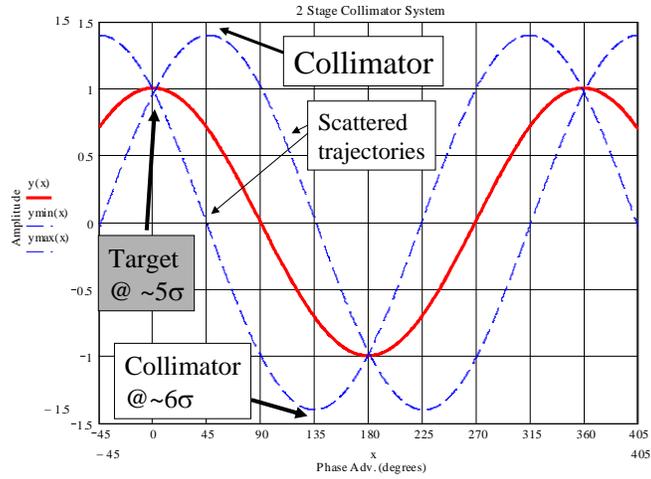

Fig.1: Placement of the target and secondary collimators to produce a 2-stage collimator system.

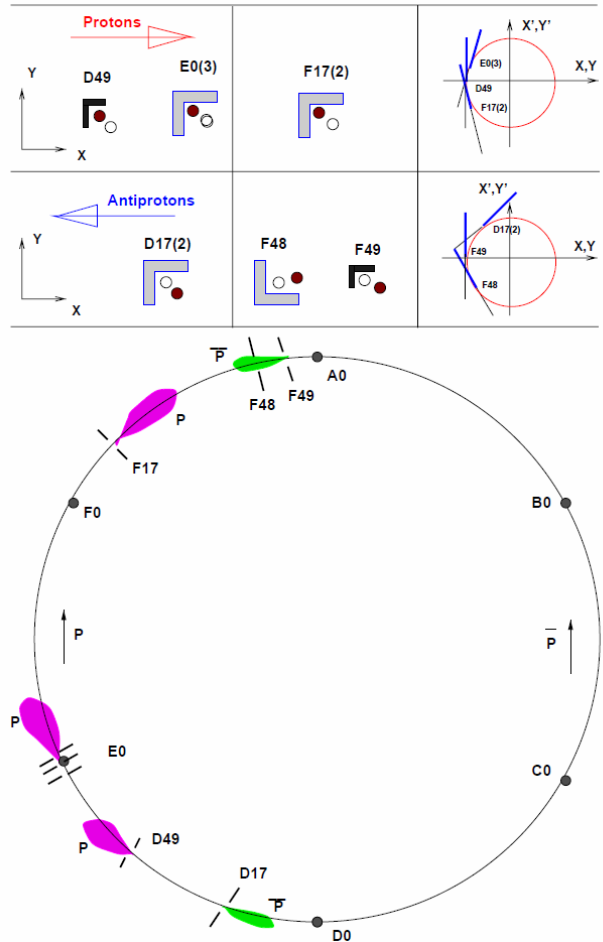

Fig.2: Tevatron Collider Run II Halo Removal Collimator Layout. CDF and D0 detectors are located at B0 and D0 respectively.

The Tevatron Collider II halo removal system requires 12 collimators of which there are 4 primary collimators or targets and 8 secondary collimators. The collimators are arranged in 4 sets: 2 proton and 2 antiproton sets and are installed around the Tevatron ring as shown in Figure 2. Placement of collimators in the Tevatron is limited to a few locations since there is limited warm space and the proton and antiproton beams are on helical orbits.

A proton primary collimator is placed at the beginning of the D17 straight section outward and up of the closed orbit (Fig. 2). It intercepts the large amplitude protons and positive off-momentum beam. Protons scattered from this collimator are represented by a vertical line in the transverse phase diagram (Fig. 2). Protons with a positive angle are intercepted by a D17(3) secondary collimator at the end of the D17 straight section. An AØ secondary proton collimator positioned outward and up of the circulating beam is intended to intercept the negative angle protons emitted from the primary collimator. A primary collimator D49 and secondary collimator F17(2) are used to deal with the protons with negative momentum deviations. Antiproton beam cleaning consists of primary collimators F49, F17(3) and secondary collimators D17(2), F48, F17(1) and EØ(2).

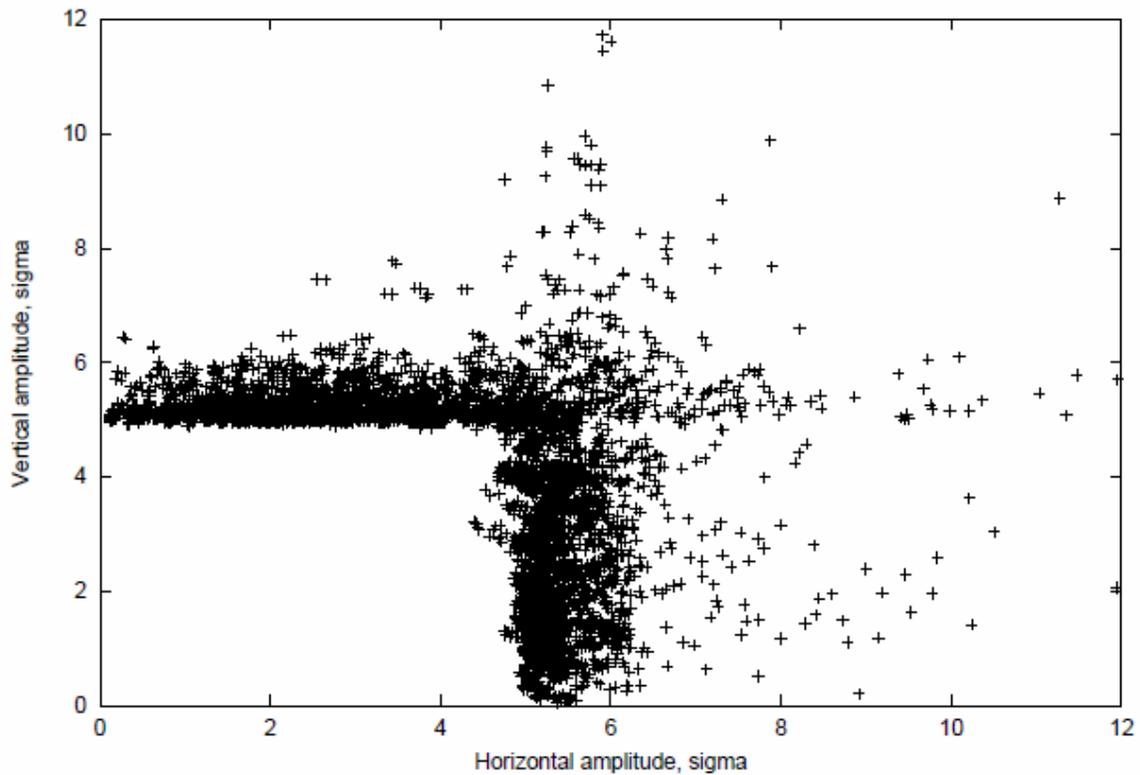

Fig.3: Proton beam halo on a secondary collimator.

Detailed STRUCT/MARS simulations assumed that halo particles first hit the primary collimator with a 1 to 3 μm impact parameter. On the next turns, the impact parameter - as a result of scattering - increases to about 0.3 mm. After the first interaction with a primary collimator, high amplitude particles are intercepted by the secondary collimators, but a large number of particles survive. Some fraction of the halo is not intercepted by a primary/ secondary collimator pair and will interact with a primary collimator on the next turns. On average, halo protons interact with the primary collimator 2.2 times. Particles with amplitudes less than $6\sigma$ are not intercepted by the secondary collimators and do survive for several tens of turns until they increase in amplitude in the next interactions with the primary collimator. The tail of the halo is extended above $6\sigma$ (Fig. 3). Large- amplitude particles, which escape from the cleaning system at the first turn, are able to circulate in the machine, before being captured by the collimators on later turns. This defines the machine geometric aperture.

The calculations [2] have shown and later measurements confirmed that the inefficiency of the Tevatron collimation system defined as a leakage of halo protons from its components is $\sim 10^{-3}$. At the same time, the most critical function of the system in Run II has been identified as reduction of background rates in the collider experiments. Beam losses at BØ and DØ depend strongly on the secondary collimator offset with respect to the primary collimators. It has been found [7] that a part of the accelerator-related backgrounds in the DØ and CDF detectors is originated from the beam halo loss in the inner triplet region. Studies [8] have revealed that the beam loss is mainly due to beam-gas elastic scattering in the regions between an inner triplet and the nearest IP secondary collimator. This process obviously can increase the background rates. In addition to the optically small aperture at $\beta_{max}$ locations, the aperture restrictions in this area are the DØ forward detector's Roman pots placed at $8\sigma$ and the BØ Roman pots placed at $10\sigma$ at the entrance and exit of the beam separators. (The Roman pot systems was removed in the middle of the Collider Run II). Thus, for the collider detectors, the above-defined inefficiency is not the whole story. The more appropriate definition of collimation inefficiency would be a ratio of backgrounds in the detectors with collimation to that without collimation. For the Tevatron Run II it is calculated as $6.7 \times 10^{-3}$, or a factor of 150 reduction of losses. The corresponding measurements are described below.

### 1.3 Operational Experience

The collimator hardware consists of a Motorola VME 162 processor and Advanced Controls System Corp. Step/Pac stepping motor drivers that interface to the VME processor [9]. LVDT's (linear voltage differential transformers) are used to read collimator positions.  Figure 4 is the block diagram for the hardware controls for a single collimator.

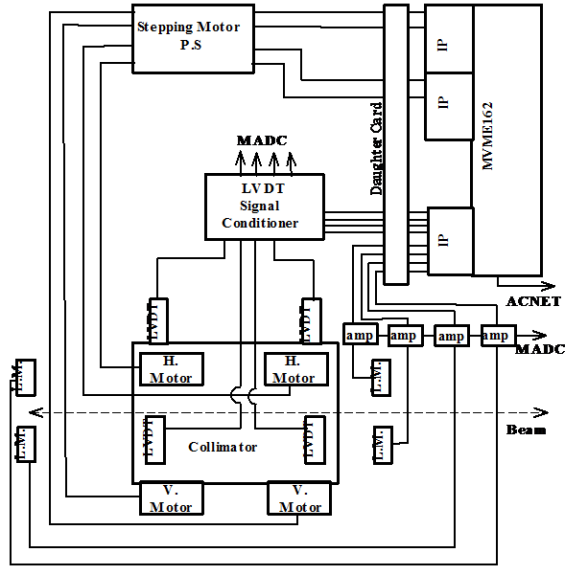

Fig.4 : Block diagram of the hardware controls for a single collimator.

The Collider II collimator halo removal system was designed with the capability of incorporating feedback into the motion of a collimator. The system uses two sources for feedback. The first source is feedback from a local beam loss monitor. Four standard Tevatron beam loss monitors and amplifiers are interfaced to the VME processor to provide loss monitor feedback. Two of these loss monitors are used to detect losses in the proton direction and two in the antiproton direction. Two loss monitors for each type of particle are used to provide redundant loss monitor signals in case of failure during collimator movement. The second source of feedback comes from a beam intensity signal. A Fast Bunch Integrator system [10] is used to provide beam intensity signals for both proton and antiproton beams at a 360Hz update rate. Feedback is accomplished by encoding proton and antiproton intensity signals on to the global machine data link (MDAT). The MDAT signal is decoded by each of the VME processors at a 720Hz rate.

Processing the feedback internal to the VME is accomplished by sampling the loss monitor and/or beam intensity signal periodically while the collimator is moving. The smallest step the collimator can make is 25μm. This minimum step takes 20msec to complete. A wait step occurs after the move step to provide more flexibility to timing movements. During this step, loss monitor signals and/or beam intensity signals are sampled every 4msec and are compared to a loss limit value or beam intensity percentage to remove value to decide if the collimator is to be halted for the next step. Each collimator VME front end has 17 parameters that the user can change to specify details about feedback processing.

The halo removal system also utilizes software that allows global coordination of all 12 local collimator VME front ends. This global coordination software is called an open access client (OAC). An OAC is a central process that runs on a central computer and has controls hooks into the main Tevatron sequencer software [11]. The OAC employs a finite state machine that is configurable by the user to preprogram one or many collimators to complete a task on a transition of a state. For example, on the state "Goto injection positions" all collimator front ends are preprogrammed with local parameters that define their out of beam positions. The OAC owns a configurable matrix of states versus collimators and the user specifies which collimators are to move when the state is transitioned. Once the state is transitioned, all collimators will be moved back to their injection positions. There are currently 11 defined collimator states with names like: Goto injection Positions, Begin Halo Removal Scraping and Retract Proton Collimators. There is one special collimator state which is "Global Collimator Abort". A transition of this state will stop all 12 collimators immediately.

The halo removal process is conducted in the Tevatron at the flattop energy of 980 GeV after the proton and antiproton beams have been brought into collisions. This process is initiated by the Tevatron sequencer software. There are four sub-sequence operations that are necessary in order to complete halo removal:

1. *Move Collimators to Initial Positions*: This sub-sequence moves all the collimators at 1.25mm/sec into the beam to the location "half way" to the beam. The motivation of this sub-sequence is to speed up the process.

2. *Intermediate Halo Removal*: Here each set (proton and antiproton) of collimators and targets are moved together under beam loss monitor feedback until a small loss is detected and all collimator in the set stop. This sub-sequence is also preformed in order to reduce the total amount of time the halo removal process takes.

3. *Perform Halo Removal*: Each secondary collimator and target is moved serially into the beam. Secondary collimators are moved under loss monitor feedback with a step size of 0.025mm until they reach the edge of the beam to shadow the losses by the primary collimator. After all secondary collimators are placed next to the beam, each target is moved under loss monitor and beam intensity feedback until 0.4% of each beam (proton and antiproton) is removed.

4. *Retract Collimators For Store*: After targets and secondary collimators have reached their final assignment, they are retracted approximately 1mm. This is the position they remain at for the duration of the store. This roughly leaves the targets and secondary collimators at the 5 and 6σ points as specified by the system design. The halo removal system is a necessary and integral part of Tevatron Collider operations. The halo removal system is completely automated and benefits operations with ease of use. The entire process takes as little as 7 min. Figure 5 presents loss rates during the process of beam collimation early in store #8709 (May 2011).

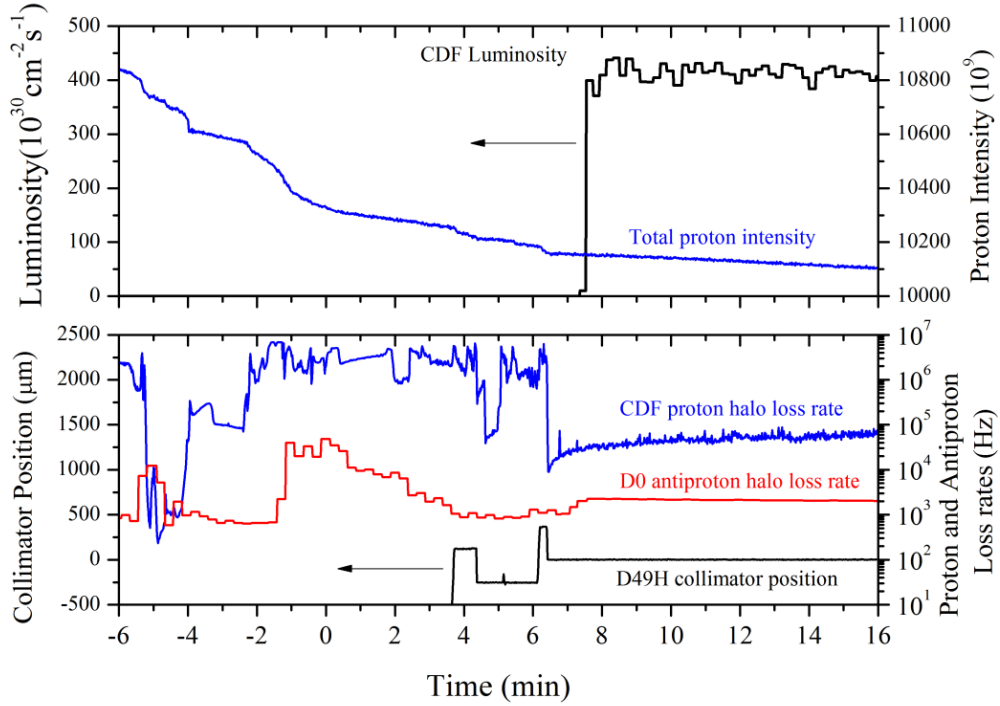

Fig. 5 : Collimation process early in store #8709 (May 2011). Upper plot shows proton beam intensity and CDF luminosity. Bottom plot shows proton halo loss rate as measured by CDF detector, D0 antiproton loss rate and horizontal position of one of the collimators (D49H). Zero time corresponds to the moment when two beams are brought to collisions.

A zero time in Fig. 5 corresponds to the moment when two 980-GeV beams are brought to collisions. Over the next six and a half minutes collimators sequentially approach the beams and scrape them. The horizontal position of one of the collimators (D49H) is shown in the bottom plot. One can see that the collimator moved very close to the beam twice – that is intentionally done to repeat the scraping procedure and guarantee lower loss rates afterward. The proton beam intensity on the upper plot shows a number of small drops due to the scraping. The bottom plot shows proton halo loss rate as measured by the CDF detector and the antiproton loss rate measured by the D0 detector. After the scraping is over – at about 7 min - the CDF detector luminosity monitor starts operation and reports maximum luminosity of about $430 \times 10^{30}$ $cm^{-2}s^{-1}$. Note that compared to the first moments after the collisions are initiated, during the luminosity operation the proton halo loss rate drops by factor of 100 from 2-3 MHz to 20-30 kHz, while antiproton rate is down by a factor of 4-5 (from 8 to 10 kHz to 2kHz).

The merit of using halo removal efficiency is to simply record the proton and antiproton halo losses at CDF and D0 IP's before halo removal divided by the same losses recorded at the completion of halo removal. Table 1 presents statistics of the reduction in the losses averaged over 100 stores in January-May 2011. One can see that the biggest reduction is seen in the CDF proton rate (over a factor of 100). The reduction in the D0 proton halo loss is relatively small, that can be attributed to the fact that for the proton direction the CDF Interaction Point acts as an addition collimator to reduce proton halo losses at D0.

TABLE 1. Merit of halo removal efficiency (2010-2011)

| Halo Loss Counter at CDF or D0 IP | Factor of reduction of halo losses after halo removal |
|---|---|
| CDF proton halo loss | 112 |
| CDF anti proton halo loss | 80 |
| D0 proton halo loss | 13 |
| D0 antiproton halo loss | 19 |

**1.4 Abort Gap Particle Removal**

Since early Run II, the electron lens [12] installed at the F48 location in the Tevatron (TEL-1) has been routinely used for removing unwanted beam particles from the abort gaps between bunch trains [13]. Particles not captured by the Tevatron radiofrequency system pose a threat because they can quench the superconducting magnets during acceleration or in case of beam aborts. Coalescing in the Main Injector typically leaves a few percent of the beam particles outside the RF buckets. These particles are transferred together with the main bunches. In addition, single intrabeam scattering (the Touschek effect), diffusion due to multiple intrabeam scattering (IBS), and phase and amplitude noise of the RF voltage, drive particles out of the RF buckets. This is exacerbated by the fact that after coalescing and injection, 95% of the particles cover almost the entire RF bucket area. The uncaptured beam is lost at the very beginning of the Tevatron energy ramp. These particles are not synchronized with the accelerating system, so they do not gain energy and quickly ($< 1$ s) spiral radially into the closest horizontal aperture. If the number of particles in the uncaptured beam is too large, the corresponding energy deposition results in a quench (loss of superconductivity) of the superconducting magnets and, consequently, terminates the high-energy physics store. At the injection energy, an instantaneous loss of uncaptured beam equal to 3-7% of the total intensity can lead to a quench depending on the spatial distribution of the losses around the machine circumference.

At top energy, uncaptured beam generation is mostly due to IBS and RF noise, whereas infrequent occurrences of longitudinal instabilities or trips of the RF power amplifiers can cause large spills of particles. Uncaptured beam particles are outside of the RF buckets, and therefore, move longitudinally relative to the main bunches. Contrary to the situation at the injection energy of 150 GeV, when synchrotron radiation (SR) losses are practically negligible, 980 GeV protons and antiprotons lose about 9 eV/turn due to the SR. For uncaptured beam particles, this energy loss is not being replenished by the RF system, so they slowly spiral radially inward and die on the collimators, which determine the tightest aperture in the Tevatron during collisions. The typical time for an uncaptured particle to reach the collimator is about 20 minutes.

To operate the TEL as the abort gap beam remover, the electron beam pulse is synchronized to the abort gap and positioned within a few millimeters of the proton beam orbit. Electric and magnetic forces due to the electron space charge produce a radial kick on the 980-GeV protons of the order of 0.1 μrad depending on electron current and separation. When the pulsing frequency of TEL-1 is near the proton beam resonant frequency, this beam-beam kick resonantly excites the betatron oscillations of the beam particles. The TEL smoothly removes the uncaptured beam from the abort gap within minutes and reduces the abort gap population by more than an order of magnitude, as demonstrated in Figure 6 from Ref. [13].

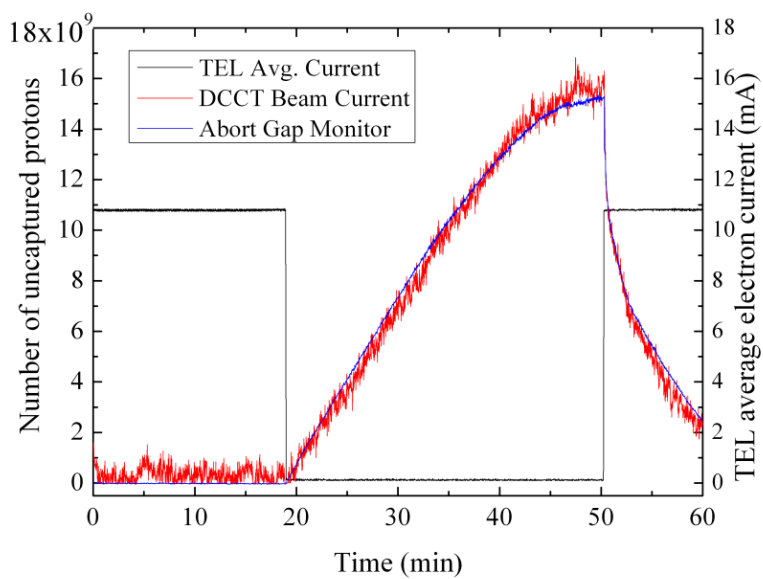

Fig. 6: Uncaptured beam accumulation when the electron current is turned off at $t = 20$ min, and subsequent removal by TEL-1. The black line represents the average electron current of the TEL; the red line is the uncaptured beam estimated from the DCCT measurement; the blue line is uncaptured beam measured by the abort-gap monitor.

At injection energy, the synchrotron radiation of protons is negligible, so the TEL is the only means to control uncaptured beam. As noted above, one of the TELs is used routinely in the Tevatron operation for the purpose of uncaptured beam removal at 150 GeV and 980 GeV. In 2007, the typical antiproton intensity increased to about a third of the proton intensity, and therefore the antiproton uncaptured beam accumulation started to pose an operational threat. An antiproton abort gap monitor (AGM), similar to the proton one, has been built and installed. By proper placement of the TEL electron beam between the proton beam and the antiproton beam, we are able to remove effectively both uncaptured protons and uncaptured antiprotons. In addition, we have explored the effectiveness of the uncaptured beam removal at several resonant excitation frequencies. For that, we have pulsed the TEL every 2nd, 3rd, 4th, 5th, 6th and 7th turn. Reduction of the uncaptured beam intensity was observed at all of them, though usually the most effective was every 7th turn pulsing when the Tevatron betatron tunes were close (slightly above) to $Q_{x,y}=4/7=0.571$ or every 6th turn pulsing when tunes were closer to $Q_{x,y}=7/12=0.583$.

## 2. New Collimation Methods

### 2.1 Bent Crystal Collimation

Since the original suggestion of bent crystal channeling [14] there has been interest in exploiting the technique for accelerator extraction [15] and, later, collimation [16]. TeV-scale channeling extraction was observed for the first time in a 900 GeV study at the Fermilab Tevatron during Collider Run I [17]. The experiment, Fermilab E853, demonstrated that useful beams can be extracted from a superconducting accelerator during high luminosity collider operations without unduly affecting the background at the collider detectors. Multipass extraction was found to increase the efficiency of the process significantly. The beam extraction efficiency was about 25%. Studies of time dependent effects found that the turn-to-turn structure was governed mainly by accelerator beam dynamics. Based on the results of the E853 experiment, it was concluded that it is feasible to construct a parasitic 5–10 MHz proton beam from the Tevatron collider [18].

The Tevatron Run II beam collimation approach has been to use a two-stage system in which a primary collimator is employed to diffuse halo particles, therefore, increasing their betatron oscillation amplitudes and impact parameters on secondary collimators (see preceding section). A bent crystal can coherently direct channeled halo particles deeper into a nearby secondary absorber. This approach has the potential of reducing beam losses in critical locations and radiation loads to the downstream superconducting magnets as was shown in the studies [19] for the Tevatron.

There are several processes which can take place during the passage of protons through the crystals: a) amorphous scattering of the primary beam; b) channeling; c) dechanneling due to scattering in the bulk of the crystal; d) "volume reflection" off the bent planes; and e) "volume capture" of initailly unchanneled particles into the channeling regime after scattering inside the crystal. A particle can be captured in the channeling regime, oscillating between two neighboring planes if it enters within crystal's angular acceptance of:

$$\theta < \theta_c = \sqrt{\frac{2U_0}{pc}} \quad (1).$$

where $p$ is the particle momentum and $U_0$ is the crystal's planar potential well depth. The critical angle $\theta_c$ is calculated to be about 7 μrad for 980 GeV/c protons in a (110) silicon crystal. When the crystal is bent, particles still can be channeled (and thus deflected) if the bend radius $R$ is greater than a critical value $R_c=pv/eE_m$, where $E_m$ is the maximum strength of the electric field in the channel, about 6 GV/cm for a (110) silicon crystal, yielding $R_c \approx 1.6$ m for 980 GeV/c protons. Bending of the crystal decreases the critical channeling angle, the capture probability of particles into the channeling regime and the dechanneling length [15]. If the particle momentum is not within the critical angle but has a tangency point with the bent planes within the crystal volume, almost all particles are deflected to the opposite direction with respect to the crystal bending. The effect is called volume reflection (VR) [15] and it has a very wide angular acceptance equal to the crystal bend angle (characteristically of the order of hundreds of microradians compared to several microradians of the channeling acceptance). The drawback of the volume reflection regime is that the deflection angle is small, approximately $(1.5-2) \times \theta_c$. However, this can be overcome by using a sequence of several precisely aligned bent crystals, so the total deflection angle is proportionally larger.

In the Tevatron beam crystal collimation experiment T980 [20]-[23] both single crystals (for vertical and horizontal deflection) and multi-strip crystal assemblies (for vertical multiple VR) have been used. Collimation of circulating beams is very different from bent crystal experiments with extracted beams [24] because of smaller initial "impact parameters" and the possibility of interplay of different effects. Figure 7 a) shows a schematic of the T-980 experimental layout. During normal Tevatron operations, a 5-mm tungsten target scatters the proton beam halo into a 1.5-m long stainless steel secondary collimator E03, 50 m downstream of the target. For the bent crystal experiments, a goniometer containing single or multi-strip bent crystals is installed 23.7 m upstream of the E03 collimator. Scintillation counter telescopes detect secondary particles from protons interacting with the target and E03 collimator. An ionization chamber (beam loss monitor LE033) also detects secondary particles scattered from E03. A PIN diode telescope detects the secondaries scattered from the bent crystal. Under the above configuration, channeled beam is signaled by a reduction of the rate in the PIN telescope with attendant increases in the rates of the LE033 and E1 counters.

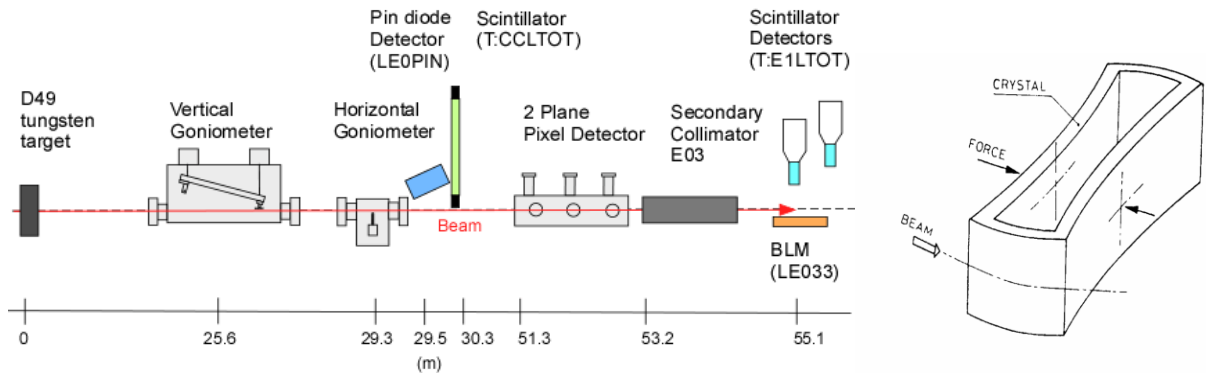

Fig.7: a) (left) General layout of T980 at E0, the straight section used for the crystal collimation test; b) (right) PNPI "O" geometry crystal used at RHIC and Fermilab. The length along the beam is 5 mm.

A modified BNL goniometer assembly [25] and an O-shaped 5-mm silicon crystal with a bending angle of 0.44 mrad were originally installed in the Tevatron downstream of the horizontal primary collimator in the fall of 2004. The crystal was set at $5.5\sigma_b \approx 2.5$ mm from the beam and aligned in the halo by varying the crystal angle in steps of several μrad. The interaction probability in the 5 mm long crystal was monitored by the PIN diode. The LE033 BLM readings are plotted as a function of the crystal angle in Fig. 8 (left). A channeling dip is present at zero angle to the crystal's plane with a width of $22 \pm 4$ μrad (rms). The width of the channeling dip is a convolution of the beam divergence, the channeling critical angle, multipass channeling effects and possible crystal distortions. It is difficult to do a deconvolution of the crystal angular scan to get the critical angle. However, the distribution is consistent with the beam divergence and measured channeling critical angle at 980 GeV of about 7 μrad, very close to the calculated one. At the bottom of the dip the LE033C signal is 22% of the signal at a random angular setting. This depth is a measure of the channeling efficiency and gives a channeling efficiency of $\eta_c = 78 \pm 12\%$ including the effects of multiple passes. A shoulder extends $460 \pm 20$ μrad to the right of the channeling dip. This shoulder width is close to the expected magnitude of the crystal bend. The shoulder is a coherent crystal effect acting over the whole arc of the crystal bend due to volume reflection. Like channeling, volume reflection will diminish nuclear interactions and thereby decrease the LE033 rate. The whole-arc efficiency, $\eta_r$, was $52 \pm 12\%$. As shown in Ref. [20], the results of Biryukov's CATCH simulation [26] for this case are in a very good agreement with the measurements, without any free parameters in the simulation except average counting rate. Most impressive, while using the bent crystal at the channeling angle instead of a tungsten primary collimator, the CDF beam losses at the opposite side of the ring dropped down by a factor of two, in a good agreement with predictions [21].

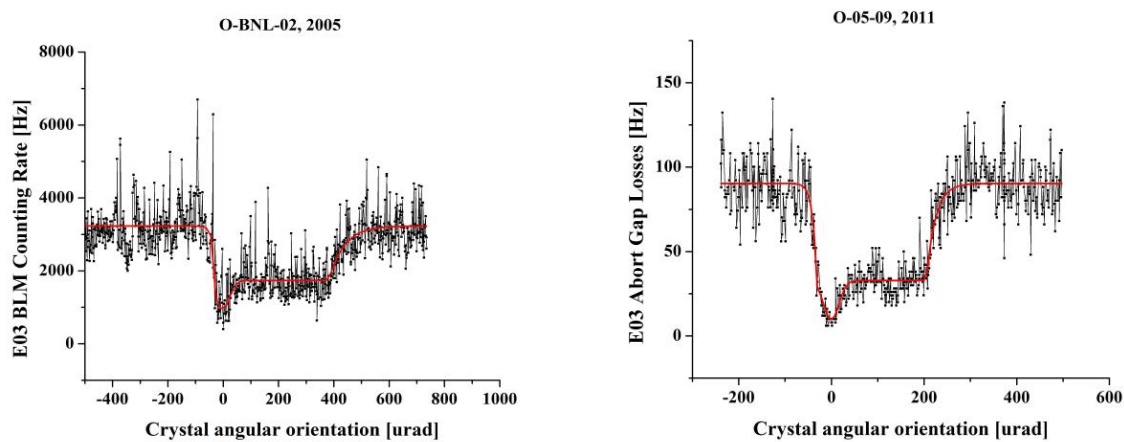

Fig. 8: Crystal angle scans for two O-shaped crystals of 2005 (left) and 2011 (right). The red curves represent the fits to the data with the Gaussian function for the channeling peak and plateau for the volume reflection region.

In 2009 the 0.44-mrad bend O-shaped crystal in the horizontal goniometer was replaced with the new 0.36-mrad O-shaped one with negative 0.12-mrad miscut angle built by IHEP (Protvino). Results of its angular scan are shown in Fig. 8 (right) for the CCLABT loss rate. A new vertical "push-pull"-type goniometer was installed 4-m upstream, housing two crystals - the multiple (eight) strip crystal from IHEP and the old 0.44 mrad O-shaped crystal, so that there were crystals for collimation in both vertical and horizontal planes. Since then crystal collimation has been routinely employed during many collider stores, and additional beam instrumentation has been added. Fast automatic insertion of the crystals has been implemented. A vertical multi-strip crystal system has been successfully tested and both multiple-VR beam at the E03 collimator and channeled beam at the F17 collimator some 1 km downstream of the E03 have been observed. A reduction of ring wide losses was reproducibly obtained along with local loss effects on the collimator due to crystal channeling and VR. The first ever study of two plane crystal collimation was also conducted.

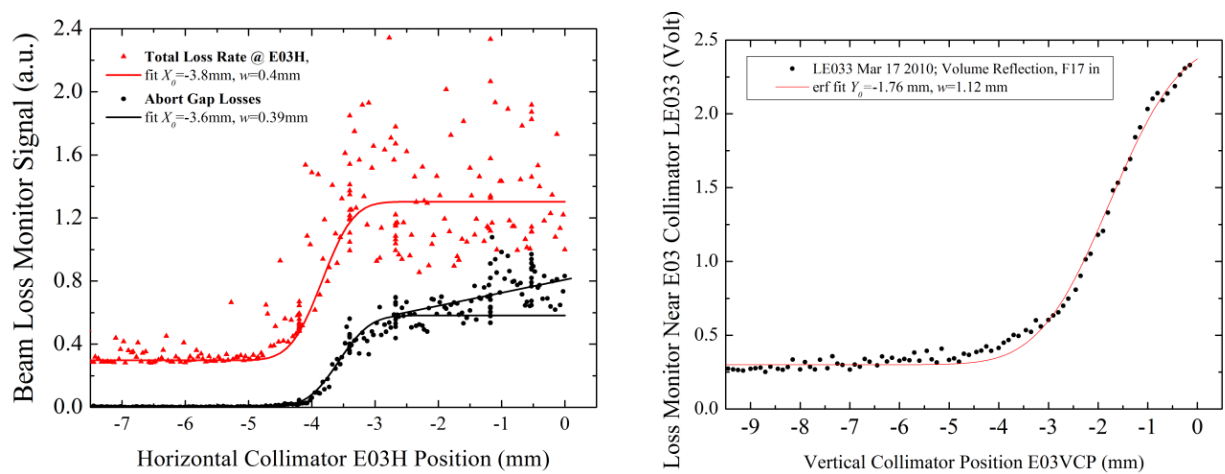

Figure 9: Collimator scan with crystal set at: a) (left) the channeling angle; b) (right) Collimator scan with 8-strip vertical crystal set at the VR angle. Solid lines are "erfc" fits of the data.

To measure the deflection of the channeled (or VR) particles once the crystal angle is set to the channeling (or VR) peak the position of an appropriate collimator can be slowly scanned, starting from a completely retracted position and moving toward the beam edge. An example of such a scan is shown in Fig.9a for horizontally deflected channeled protons at the E03H collimator. The curves show the total measured loss rate (red dots) as well as the counting rate synchronized to the abort gaps only (black dots). There are three distinct regions: a) a region of negligible losses, where the collimator does not intercept any beam; b) a steep increase in the losses, where the collimator intercepts the channeled beam; c) a region where the losses increase slowly: the collimator is additionally intercepting de-channeled and amorphous scattered particles. Both abort gap and total loss signals show a small deflection angle of (3.6-3.8)mm/24 m=150-160 μrad instead of the expected 360 μrad. Such a difference can either be attributed to the effect of the "miscut angle" [21] or be due to non-ideal crystal surface that becomes important at the very small impact parameters. The angular spread in the channeled beam is about of 0.4 mm/24 m=17 μrad rms that is larger than the channeling acceptance of $2\times\theta_c$=13.4 μrad. A similar scan of the VR beam made with the E03 vertical collimator - see Fig.9b - shows the beam at 1.76mm/28m=63 μrad, i.e., approximately where it is supposed to be, and about 40 μrad rms wide [22].

**2.2 Hollow Electron Beam Collimation**

The hollow electron beam collimator is a novel concept of controlled halo removal for intense high-energy hadron beams in storage rings and colliders [27], [28]. It is based on the interaction of the circulating beam with a 5-keV, magnetically confined, pulsed hollow electron beam in a 2-m-long section of the ring. The electrons enclose the circulating beam, kicking halo particles transversely and leaving the beam core unperturbed (Figure 10a and 10b). By acting as a tunable diffusion enhancer and not as a hard aperture limitation, the hollow electron beam collimator extends conventional collimation systems beyond the intensity limits imposed by tolerable losses. The concept was tested experimentally at the Tevatron between October 2010 and September 2011. It represents a promising option for scraping high-power beams in the Large Hadron Collider.

In high-power hadron machines, conventional two-stage collimation systems offer robust shielding of sensitive components and are very efficient in reducing beam-related backgrounds at the experiments. However, they have limitations. The minimum distance between the collimator and the beam axis is limited by instantaneous loss rates (especially as jaws are moved inward), radiation damage, and by the electromagnetic impedance of the device. Moreover, beam jitter, caused by ground motion and other vibrations and only partially mitigated by active orbit feedback, can cause periodic bursts of losses at aperture restrictions. The hollow electron beam collimator (HEBC) addresses these limitations, emerging as a viable complement to conventional systems.

In the Tevatron electron lenses, the electron beam is generated by a pulsed 5-kV electron gun and transported with strong axial magnetic fields. Its size in the interaction region is controlled by varying the ratio between the magnetic fields in the main solenoid and in the gun solenoid. Halo particles experience nonlinear transverse kicks and are driven towards the collimators. If the hollow current distribution is axially symmetric there are no electric or magnetic fields inside, and the beam core is unperturbed. A magnetically confined electron beam is stiff, and experiments with the electron lenses showed that it can be placed very close to, and even overlap with the circulating beam. Another advantage is that, contrary to conventional systems, no nuclear breakup is generated in the case of ion collimation. In a setup similar to that of the Tevatron electron lenses, with a peak current of 1 A, an overlap length of 2 m, and a hole radius of 3 mm, the corresponding radial kick is 0.3 μrad for 980-GeV antiprotons. The intensity of the transverse kicks is small and tunable: the device acts more like a soft scraper or a diffusion enhancer, rather than a hard aperture limitation. Because the kicks are not random in space or time, resonant excitation is possible if faster removal is desired.

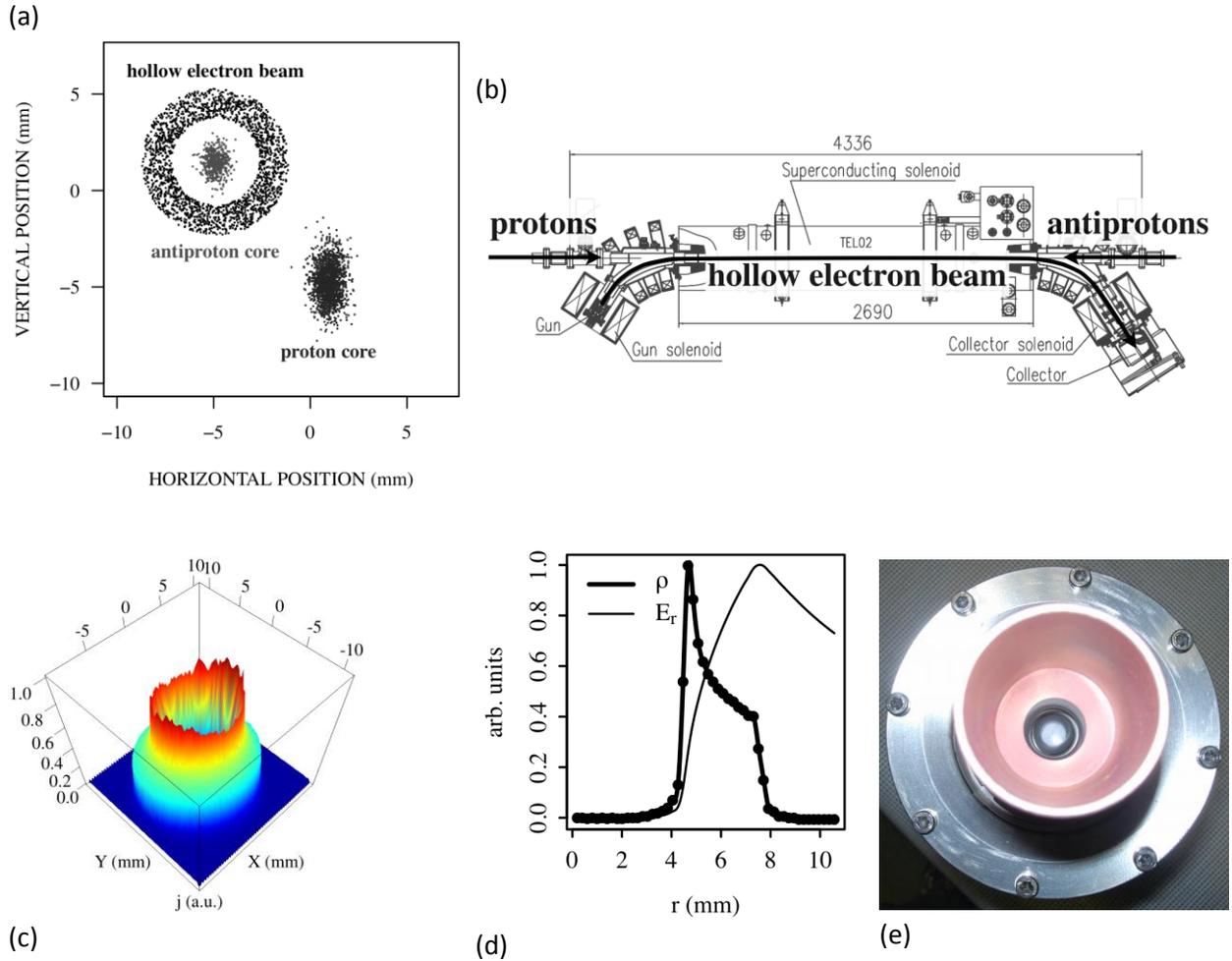

Figure 10: Tevatron hollow electron beam collimator: (a) transverse beam profiles; (b) top view of the system layout; (c) measured current profile; (d) measured charge density $\rho$ and calculated radial electric field $E_r$; (e) photograph of the electron gun.

Analytical expressions for the current distribution were used to estimate the effectiveness of the HEBC on a proton beam. They were included in tracking codes such as STRUCT [3], Lifetrac [29], and SixTrack [30] to follow core and halo particles as they propagate in the machine lattice. These codes are complementary in their treatment of apertures, field nonlinearities, and beam-beam interactions. Preliminary simulations suggested that effects would be observable and that measurements would be compatible with normal collider operations [31], [32].

A 15-mm-diameter hollow electron gun was designed and built (Figure 10c, 10d, and 10e) [33], [34]. It is based on a tungsten dispenser cathode with a 9-mm-diameter hole bored through the axis of its convex surface. The peak current delivered by this gun is 1.1 A at 5 kV. The current density profile was measured on a test stand by recording the current through a pinhole in the collector while changing the position of the beam in small steps. The gun was installed in one of the Tevatron electron lenses (TEL-2) in August 2010. The pulsed electron beam could be synchronized with practically any bunch or group of bunches.

The behavior of the device and the response of the circulating beams were measured for different beam currents, relative alignments, hole sizes, pulsing patterns, and collimator system configurations. Preliminary results were presented in Refs. [35], [36]. Here, we discuss a few representative experiments illustrating the main effects of the electron beam acting on antiproton bunches. Antiprotons were chosen for two main reasons: their smaller transverse emittances (achieved by stochastic and electron cooling in the Antiproton Source accelerators) made it possible to probe a wider range of confining fields and hole sizes; and the betatron phase advance between the electron lens and the absorbers was more favorable for antiproton collimation.

The particle removal rate was measured by comparing bunches affected by the electron lens with other control bunches. In the experiment described in Figure 11, the electron lens was aligned and synchronized with the second antiproton bunch train, and then turned on and off several times at the end of a collider store. The electron beam current was about 0.4 A and the radius of the hole was varied between $6\sigma_y$ and $3.5\sigma_y$, $\sigma_y = 0.57$ mm being the vertical root-mean-square beam size. The black trace is the electron-lens current. One can clearly see the smooth scraping effect. The corresponding removal rates are of a few percent per hour.

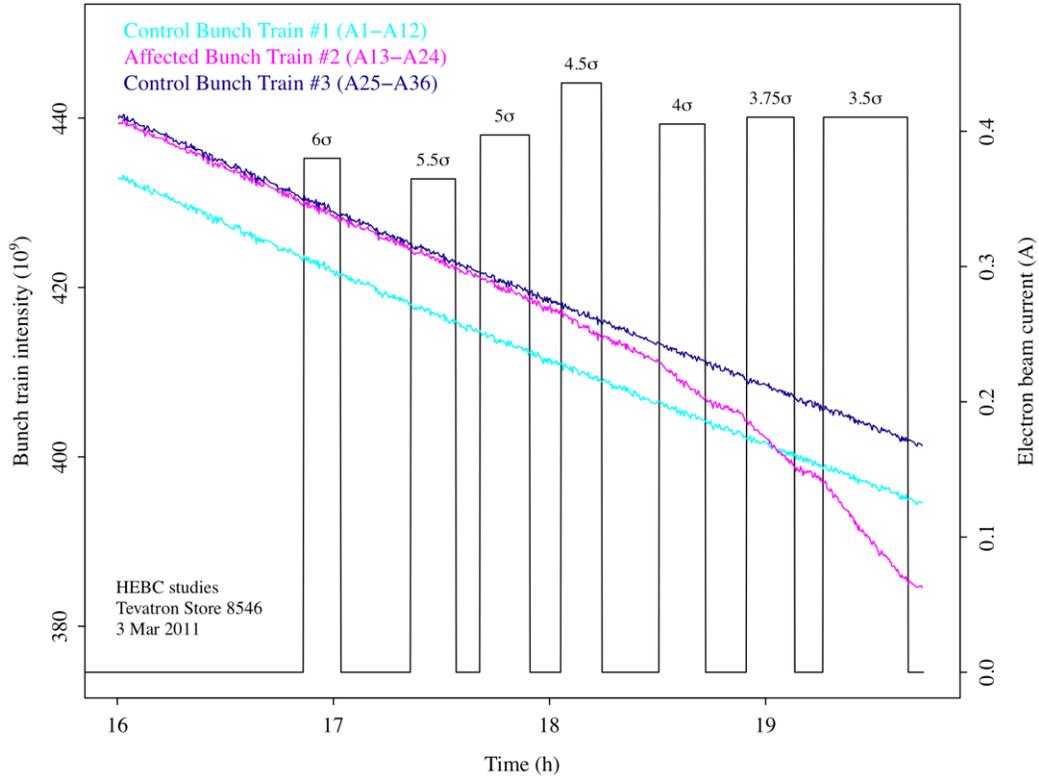

Figure 11: Scraping effect of the hollow electron beam acting on one antiproton bunch train (magenta) at the end of a collider store. The intensities of the two control trains are shown in cyan and blue. The electron beam current (black trace) was turned on and off several times with different values of the hole radius, from $6\sigma_y$ to $3.5\sigma_y$.

Whether there are any adverse effects on the core of the circulating beam is a concern. The overlap region is not a perfect hollow cylinder, due to asymmetries in gun emission, evolution under space charge of the hollow profile, and the bends in the transport system. The problem was approached from several points of view. First, one can see from Figure 11 that no decrease in intensity was observed with large hole sizes, when the hollow beam was shadowed by the primary collimators. This implies that the circulating beam was not significantly affected by the hollow electron beam surrounding it, and that the effect on beam intensity of residual fields near the axis was negligible. Second, we observed no difference in emittance growth for the affected bunches: if there was emittance growth produced by the electron beam, it was much smaller than that driven by the other two main factors, namely intrabeam scattering and beam-beam interactions. The effect of halo removal can also be observed by comparing beam scraping with the corresponding decrease in luminosity. Luminosity is proportional to the product of antiproton and proton populations, and inversely proportional to the overlap area. If antiprotons are removed uniformly and the other factors are unchanged, luminosity should decrease by the same relative amount. If the hollow beam causes emittance growth or proton loss, luminosity should decrease even more. A smaller relative change in luminosity was observed, which is a clear indication of halo scraping. Also, the ratio between luminosity decay rates and intensity decay rates increased with decreasing hole size. Finally, one can attempt to directly measure the particle removal rate as a function of amplitude. This was done with collimator scans. A primary antiproton collimator was moved vertically in 50-micron steps towards the beam axis. All other collimators were retracted. The corresponding beam losses and decay rates were recorded. Particles were removed from the affected bunch train, but as soon as the primary collimator shadowed the electron beam, eliminating the halo at those amplitudes, the relative intensity decay rate of the affected bunch train went back to the value it had when the lens was off. Even with a hole size of $3.5\,\sigma_y$, the effects of residual fields on the core appeared to be negligible. The time evolution of losses during a collimator scan was also used to measure changes in diffusion rate as a function of amplitude, using an extended version of the technique presented in Refs. [6, 37].

Another observation was that the hollow electron lens mitigated the effects of beam jitter. In the Tevatron, beams oscillate coherently at low frequencies (from sub-hertz to a few hertz) with amplitudes of a few tens of microns, due to mechanical vibrations and ground motion [38]. This causes periodic bursts of losses at aperture restrictions, with peaks exceeding a few times the average loss rate. When the collimators are moved inward, these loss spikes can cause quenches in the superconducting magnets or damage electronic components. In March 2011, to measure the loss spikes and the effects of the hollow electron beam, scintillator paddles were installed downstream of one of the antiproton secondary collimators (F48). These loss monitors could be gated to individual bunch trains. It was observed that losses from the two control trains were completely correlated, and that – as shown in Fig. 12 (top) - their frequency spectra exhibited strong peaks at 0.39 Hz and its harmonics (corresponding to the acceleration cycle of the Main Injector) and at 4.6 Hz (mechanical vibrations from the Central Helium Liquefier). As Fig. 12 (bottom) shows, the electron lens suppressed these peaks and eliminated correlations with the other trains. This can be interpreted as a reduction in the population of the beam tails, which makes the affected bunch train less sensitive to coherent beam oscillations.

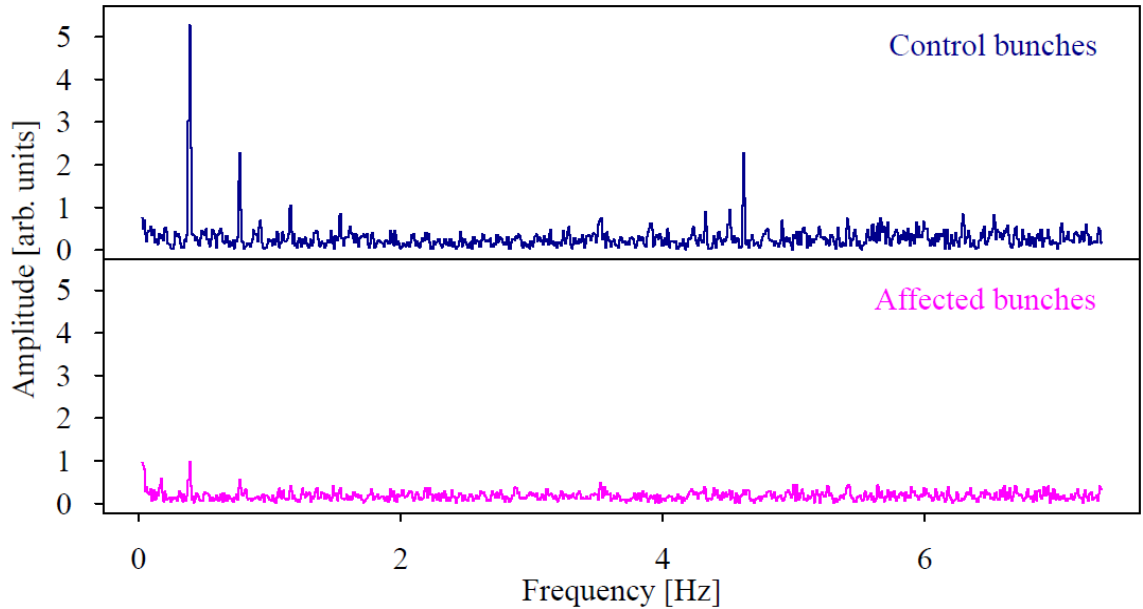

Figure 12: Frequency spectra of losses measured near the F48 antiproton collimator coming from different bunches: the control bunch trains (top) and the bunch train affected by the hollow electron beam (bottom).

Losses generated by the electron lens were mostly deposited in the collimators, with small changes at the experiments. Alignment of the beams was done manually, with a setup time of about 15 minutes. Alignment is crucial for HEBC operation, and the procedures based on the electron-lens beam-position monitors were found to be reliable in spite of the different time structure of the electron and (anti)proton pulses. No instabilities or emittance growth were observed over the course of several hours at nominal antiproton intensities ($10^{11}$ particles/bunch) and electron beam currents up to 1 A in confining fields above 1 T in the main solenoid. Most of the studies were done parasitically during regular collider stores.

Experiments at the Tevatron showed that the hollow electron beam collimator is a viable option for scraping high-power beams in storage rings and colliders. Its applicability to the LHC is under study. To make the device more versatile, larger cathodes and higher electron beam currents appear to be feasible, and experimental tests in this direction are planned.

**Acknowledgments**

We would like to thank R. Assmann, M. Convery, C. Gattuso, V. Kamerdzhiev, R. Moore, A. Romanov, and G. Saewert for many useful discussions on the subject of this report.